\documentstyle[pre,aps,preprint,epsf]{revtex}
\tightenlines
\begin{document} 

\title{\bf Linear and Nonlinear Rheology of 
Wormlike 
Micelles}
\author{
A. K. Sood, Ranjini Bandyopadhyay and Geetha Basappa
}
\address{
Department of Physics\\
Indian Institute of Science\\
Bangalore 560 012\\
India
}
\maketitle

\begin{abstract}
   Several surfactant molecules self-assemble in solution to form 
long, cylindrical, flexible wormlike micelles. These micelles can be 
entangled 
with each other leading to viscoelastic phases. The rheological 
properties of such phases are very interesting and have been the
subject of a large number of experimental and theoretical studies 
in recent years. We will report our recent work on the 
macrorheology, 
microrheology and nonlinear flow behaviour of dilute aqueous
solutions of a surfactant CTAT (Cetyltrimethylammonium Tosilate). 
This system forms elongated micelles and exhibits strong 
viscoelasticity at low concentrations ($\sim$ 0.9 wt\%) without the
addition of electrolytes. Microrheology measurements of $G(\omega)$ 
have been done using diffusing wave spectroscopy which will be 
compared with the conventional frequency sweep measurements done
using a cone and plate rheometer. The second part of the paper 
deals with the nonlinear rheology 
where the measured shear stress 
$\sigma$ is a nonmonotonic function of the  shear rate $\dot{\gamma}$.
In stress-controlled experiments, the shear stress 
shows a plateau for $\dot{\gamma}$ larger than some critical 
strain rate, similar to the earlier reports on CPyCl/NaSal 
system. Cates et al have proposed that the 
plateau is a signature of mechanical instability in the form of 
shear bands. 
We have carried out extensive experiments under controlled strain 
rate conditions, to study the time-dependence of shear stress. 
The measured time series of shear stress has been analysed in terms 
of correlation integrals and Lyapunov exponents to show unambiguously
that the behaviour is typical of low-dimensional dynamical systems. 

\end{abstract}

\section{\bf{Introduction}}

Rheology is the study of the deformation and flow of matter. Solids and fluids
exhibit different flow 
behaviours under shear. Solids store mechanical energy and are 
elastic, whereas fluids dissipate energy and are viscous. 
Complex fluids (e.g. colloids, polymers), owing to their larger length scales  
which result in low mechanical susceptibilities ($\sim 10^2$ Pa as 
compared to $\sim 10^{12}$ Pa in atomic systems), show very 
complex flow behaviour and are viscoelastic.  
The relative proportion of elastic and viscous responses
depends on the frequency of the applied stress. For example, for entangled
polymer solutions, the stress relaxation predominantly occurs by reptation dynamics 
with time scales $\tau_{rep}$. This system will be more elastic for
$\omega > \tau_{rep} ^{-1}$ whereas it will be more viscous for 
$\omega < \tau_{rep} ^{-1}$. One should also note that for soft condensed
matter, the elastic modulus under shear stress is much smaller than 
the elastic modulus under compressive stress whereas these two moduli are
nearly equal for conventional atomic systems.

Experiments on the rheology of matter involve the measurement and
prediction of its flow behaviour. The method involves the application of 
a known strain or strain rate to a sample and the 
subsequent measurement of the stress 
induced in the sample or vice versa. 
The response of a viscoelastic material to an applied stress may be 
characterized 
as linear or nonlinear depending on the magnitude of the applied
stress/strain rate. 

\subsection{\bf{Linear Rheology}}

The response of a material is linear when very small stresses (i.e. small 
compared to the spontaneous thermal fluctuations in the material) 
are applied. If a small step 
strain $\gamma$ is applied to a deformable material at time $t=t_1$,
the stress induced in the material is given by 
$\sigma_{xy} (t) = G(t,t_1) \gamma (t) $ (Fig. 1),
where $G(t,t_1)$ is the stress relaxation function. 
Here x and y define the velocity and velocity gradient directions, 
respectively. 
To linear order in $\gamma$, 
all other components of stress like $\sigma_{xx}$ are zero.
Invoking time-translational
symmetry, $G$ depends on the time
difference between $t$ and $t_{1}$ i.e. $G(t,t_1) = G(t-t_1)$. 
Exceptions to time-translational
symmetry are found to occur in glassy systems which
show aging behaviour. 
For an arbitrarily small applied strain rate 
$\dot\gamma (t)$, the stress $\sigma_{xy}(t)$ induced in the material is defined as 
\begin{equation}
\sigma _{xy}(t) = \bigl\lmoustache ^{t}_{- \infty} G(t-t^{\prime})\dot\gamma 
(t^{\prime}) 
dt^{\prime},
\end{equation}
where G(t) is known as the memory kernel for shear response. 


{\bf{(a) Oscillatory flow}}

If a strain $\gamma (t) = \gamma_0 e^{i \omega t}$ is applied to the material,
the stress developed in the sample can be out of phase with the strain 
by a phase angle $\delta$ (Fig. 2). For
a viscoelastic material, $\delta$ lies between the limits $\delta$ = 0 (for a 
Hookean solid) and $\delta$ = $\pi$/2 (for a Newtonian fluid). 
Using Equation 1, the resulting stress may be written as
\begin{equation}
\sigma _{xy}(t) = \gamma_{o}\bigl\lmoustache ^{t}_{- \infty} 
G(t-t^{\prime})i \omega e^{i \omega t^{\prime}} dt^{\prime}.
\end{equation}
Putting $ \tau = t-t^{\prime}$, it can be written as
$\sigma_{xy}(t) = \gamma_0 e^{i \omega t} G^{\star} (\omega)$, 
$G^{\star} (\omega) = G^{\prime}(\omega) + i G^{\prime \prime}(\omega)$, 
is the complex shear modulus given by 
\begin{equation}
G^\star (\omega) =i \omega \bigl\lmoustache ^{\infty}_{0} 
G(t) e^{-i \omega t} dt.
\end{equation}
The real part of 
$G^\star(\omega)$, called the storage modulus and denoted by 
$G^\prime(\omega$), gives the elastic response of the material to 
the 
applied strain. It is the ratio of the stress in phase with the 
applied strain to the strain. The viscous response, defined by 
the loss modulus $G^{\prime\prime}(\omega)$,  
is the ratio of the out-of-phase 
stress 
component to the strain.
Equation 3 implies that $G^{\prime}(\omega)$ is an even function
whereas $G^{\prime\prime}(\omega)$ is an odd function of $\omega$.

{\bf{(b) Steady shear}}

When $\dot\gamma (t)$ = constant, the stress induced in the sample
is given by $\sigma_{xy} = \dot\gamma \bigl\lmoustache ^{t}_{-\infty} 
G(t-t^{\prime}) dt^{\prime}=\eta_{o} \dot\gamma$, 
where $\eta_{o}$ is the zero shear viscosity given by
$\eta_{o} = \bigl\lmoustache ^{\infty}_{0} G(t) dt$. 
A non-zero
value of $\eta_{o}$ implies the presence of liquid-like or glassy dynamics. 

The simplest form of the response function G(t) is given by the Maxwell model,
 namely, $G(t) = G_{o} e^{-t/\tau_{M}}$, or
$G^\star(\omega) = G_{o} i \omega \tau_{M} /(1+i \omega \tau_{M})$, 
where $G_{o}$ is the elastic modulus and $\tau_{M}$ is the Maxwell 
relaxation time related to zero frequency shear viscosity 
$\eta_{o} = G_{o} \tau_{M}$. For a Newtonian fluid, 
$G_{o}\ \longrightarrow \infty ,\ \tau _{M\ }\longrightarrow 0$ 
such that $\eta_{o}$ remains constant. More generally, when there are many
relaxation times present in the system, 
$G(t)\ =\ \sum _{j}\ G_{j}\ e^{\frac{-t}{\tau _{j}}}$.
For a continuous distribution of relaxation times, specified in 
terms of $P(\tau)$, 
\begin{equation}
G(t) = G_{o} \bigl\lmoustache ^{\infty}_{0} 
P(\tau) e^{-t/\tau} d{\tau} \equiv G_{o} \mu (t) .
\end{equation}
For entangled polymer solutions, the stress relaxation occurs by
reptation dynamics with time scale $\tau_{rep}$, for which 
$\mu (t)\ =\ \sum _{n\ odd\ }{8\over n^{2}\pi ^{2}}e^{({-n^{2}t\over \tau
_{rep}})}$  

In a linear creep experiment, a step response is applied 
($\sigma_{xy} =$ 0 for t $\leq$ 0 and $\sigma_{xy} = \sigma_{o}$ 
for t $>$ 0) and strain $\gamma (t)$ is measured, which is a solution of 
\begin{equation}
\sigma_{o} = \bigl\lmoustache ^{t}_{0} G(t-t^{\prime}) 
\dot\gamma (t^{\prime}) dt^{\prime}.
\end{equation}

The linear response of a viscoelastic material to an applied 
stress may be determined by a conventional frequency sweep 
experiment done using a rotating disc or 
concentric cylinders rheometer. The rheometer applies 
very small oscillatory stresses and calculates the resultant 
strain in the sample over the desired frequency range. For an 
applied angular frequency 
$\omega$, the response that is in phase with the 
applied stress is used to calculate the storage modulus 
$G^\prime(\omega)$ while the out of phase component gives the 
loss modulus $G^{\prime\prime}(\omega)$.

\subsection{\bf{Nonlinear Rheology}}

Nonlinear rheology describes the response of a material to much larger stresses. 
As the name suggests, the strain induced in a sample varies         
nonlinearly with the applied stress in this regime. The
nonlinear behaviour of a viscoelastic material 
in steady flow experiments is characterized
by shear thinning or thickening, the presence of non-zero
yield stress and normal stress differences, flow-induced
phase transitions and the phenomenon of shear banding \cite{reh,spe,ber,fis} 
as shown schematically in Fig. 3. In nonlinear step strain experiments,
if a large enough step strain $\gamma_0$ is applied to a sample, then the 
stress induced in the sample may be expressed by 
$\sigma _{xy} (t) =\gamma_0 G_{nl}(t-t^{\prime},\gamma_0)$. 
The normal stresses under these conditions are no longer negligible
and must be taken into consideration.
For $\gamma_{o} \rightarrow$ 0, $G_{nl} \rightarrow$ G(t) as measured
in linear rheology.



Systems of giant wormlike micelles formed in 
certain surfactant solutions are known to show very unusual 
nonlinear rheology\cite{reh}. In 
steady shear, the shear stress saturates to a constant value 
above a critical strain rate $\dot\gamma$ (as shown in Fig. 3 (d))
while the first 
normal stress difference increases roughly linearly with 
shear rate \cite{spe}.
 Such behaviour is a signature of 
mechanical instability of the shear banding type\cite{spe,gra} and may 
be understood in terms of the reptation-reaction model which involves the
reversible breakage and recombination of wormlike 
micelles along with repation 
dynamics known for polymer solutions. Alternatively, the 
non-monotonicity of the flow curve has been attributed to the coexistence
of two thermodynamically stable phases (isotropic and nematic) 
in the sheared solution \cite{ber2}. 

The flow curve may be measured under conditions of controlled stress 
or strain rate, and depending on the time interval between the collection of 
data points, we can obtain  metastable or steady-state branches, 
respectively.
In stress relaxation experiments, a constant step strain rate 
is applied to 
the sample in the nonlinear
regime, 
following which  the relaxation of stress in the sample is measured 
as a function of time. Alternatively,
stress relaxation may be studied after cessation of a controlled strain
rate that had been applied to the sample for a known duration.

\section{\bf{Our Experiments on CTAT aqueous solutions}}

We have studied the linear and nonlinear rheology of dilute 
aqueous solutions of the surfactant system CTAT 
(Cetyltrimethylammonium Tosilate) at 25$^\circ$C. 
Above 
concentrations of 0.04 wt.\% , and
temperatures of 23$^\circ$C \cite{sol1}, CTAT self-assembles to form 
cylindrical worm-like micelles which get entangled at concentration 
$>$ 0.9 wt.\%. 
The lengths of these wormlike micelles depend on the 
concentrations of the surfactant and the added salt, the 
temperature and the energy of scission of the micelle. The        
energy of scission is 
the excess free energy of a pair of hemispherical end caps 
relative to the rod like region containing an equivalent number 
of 
surfactant molecules. The number density of the elongated 
micelles 
of 
length L is given by \cite{sol1,can}
\begin{equation}
C_{o}(L)\sim {1\over L^{2}}exp(-{L\over L_{avg}})  
\end{equation}
where L is expressed in monomer units and
\begin{equation}
L\sim \phi ^{0.5}exp({E_{scis}\over 2k_{B}T})
\end{equation}
where $\phi$ is the surfactant volume fraction and $E_{scis}$ is 
the energy of scission of the micelle.
In these systems, stress relaxation occurs by reptation 
with time scale $\tau_{rep}$ (the     
curvilinear diffusion of the micelle through an imaginary tube   
segment) as for conventional polymers and by the reversible scission
(breakdown and recombination of micelles with time scale $\tau_b$) 
\cite{spe}. The time scales $\tau_{rep}$ and $\tau_{b}$ 
may or may not be comparable and depend on 
the surfactant concentration, presence of counterions in the 
solution and temperature.

{\bf{(a) Macrorheology measurements}}

The frequency response of a viscoelastic material may be measured using 
a rheometer which consists of a device that can simultaneously apply a torque and
measure the resultant strain. The in-phase and 
out-of-phase responses of the material to the torque are measured to calculate 
its elastic and viscous moduli, respectively.
The instrument used by us is Rheolyst AR-1000N (T.A. 
Instruments, U.K.)
stress controlled rheometer with temperature control and software for 
strain rate control to measure the elastic and viscous responses 
of 1 wt.\% CTAT between the angular frequency range of 0.03 rad/sec and 10
rad/sec. 
The rheometer used was equipped with four strain gauge transducers capable 
of measuring the normal force with an accuracy of $10^{-4}$ N.
The measurements were made using a cone-and-plate 
geometry of cone diameter 4 cm and angle 1$^\circ$59$^{\prime\prime}$. 

The linear regime of CTAT was first ascertained by looking for a 
range of stress values where 
the magnitude of the response functions were found to be independent of the 
applied oscillatory stress. The elastic and viscous moduli and the viscosity of
CTAT 1 wt.\% at 25$^\circ$C were found to be constant for stresses between 0.05 and 0.1 Pa,
oscillating 
at a frequency of 0.1 Hz. Hence for the linear response measurement, 0.08 Pa was  
chosen as the amplitude of the 
oscillatory stress, oscillating between angular frequencies 0.03  
rad/sec and 10 rad/sec. At frequencies higher than 10 rad/sec, the 
waveform depicting the strain becomes distorted, possibly due to the 
slip between the sample and the plates. This, therefore, limits the 
measurements till 10 rad/sec.
Linear response measurements (Fig. 4 ) show that at the
lowest frequencies, CTAT behaves like a viscous material, whereas in higher 
frequency runs, the behaviour is found to be predominantly elastic. The         
crossover is found to occur at 0.45 rad/sec which corresponds to a 
relaxation time $\tau_R$ of 2.2 seconds. 
Cates et al \cite{can} have shown that for a system of wormlike micelles, 
like CPyCl/NaCl, $G^{\prime}(\omega)$ and $G^{\prime\prime}(\omega)$ 
are given by Maxwell model: 
\begin{eqnarray*}
G^{\prime}(\omega) = G_{o}{\omega}^{2}{\tau_{R}}^2/(1+{\omega}^{2}{\tau_{R}}^2),\\
G^{\prime\prime}(\omega) = G_{o}{\omega}{\tau_{R}}/(1+{\omega}^{2}{\tau_{R}}^2), 
\end{eqnarray*}
where $\tau_{R}=(\tau_{b}\tau_{rep})^{1/2}$. 
Figs 5 (a) and (b) show the least square fits of the data
to the Maxwell model giving $G_{o}$ = 2.1 Pa
and $\tau_{R}$ = 2.2 sec. 
We find that for CTAT at concentration 
1 wt.\%, the fit is very poor. Further, the Cole-Cole plot (Fig 5 (c))
corresponding to the above data shows a deviation from the semi-circular 
behaviour expected in Maxwellian systems and shows an upturn at high frequencies. 
This deviation from Maxwellian behaviour is 
possibly due to the comparable values of     
$\tau_{rep}$ and $\tau_{b}$ in this system unlike in other wormlike 
micellar systems where the differences in the time scales 
($\tau_{b} << \tau_{rep}$), lead to a 'motional averaging' effect.
We have tried the Doi-Edward model where\cite{sol2} 
\begin{eqnarray*}
G^{\prime}(\omega) = G_{o}\sum_{p\ odd}({\omega}{\tau_{D}}/p)^2/(1+({\omega}{\tau_{D}}/p)^2)\\ 
G^{\prime\prime}(\omega) = G_{o}\sum_{p\ odd}({\omega}{\tau_{D}}/p)/(1+({\omega}{\tau_{D}}/p)^2)
\end{eqnarray*}
Here also, the fit with p = 1 and 3 is poor as shown in Fig. 6. The Doi-Edward model gives
$G_{o}\sim$ 3 Pa and $\tau_D \sim$ 1 sec. We also find that the Hess 
model which is given by\cite{sol2} 
\begin{eqnarray*}
G^{\prime}(\omega) = ((\eta_{o}-\eta_{\infty})/ \tau_{\epsilon})
{\omega}^{2}{\tau_{\epsilon}}^2/(1+{\omega}^{2}{\tau_{\epsilon}}^2)\\
G^{\prime\prime}(\omega) = ((\eta_{o}-\eta_{\infty})/\tau_{\epsilon}) 
{\omega}{\tau_{\epsilon}}/(1+{\omega}^{2}{\tau_{\epsilon}}^2)
\end{eqnarray*}
does not fit our data over the entire frequency range. In this model, 
 $\eta_{\infty}$ is the high frequency shear
viscosity and $\tau_{\epsilon}$ 
is a characteristic relaxation time of the system. It is likely that 
$G^{\prime}(\omega)$ and $G^{\prime\prime}(\omega)$ can be fitted with a 
model calculated in low and high frequency regions separately, as 
is usually done in polymer 
literature\cite{doi}.
Linear response measurements were also made for CTAT 
1.9wt.\% and CTAT 5wt.\%. Interestingly, 
CTAT 1.9wt.\% shows an anomalously large relaxation 
time, whereas CTAT 5 wt.\% is found to exhibit Maxwellian behaviour, as
also seen in recent studies \cite{sol2}.

{\bf {(b) Microrheology measurements}}

In recent years, micro-rheological techniques have been developed 
in addition to macroscopic rheometry measurements using rotating 
disc or concentric cylinder rheometer. 
The basic idea behind microrheology is the 
tracking or manipulation of sub-micrometer particles immersed in the
viscoelastic medium to be studied. Using magnetic beads as the probe 
particles, which can be manipulated by magnetic field gradients, the 
viscoelastic properties of F-actin networks \cite{zan} and the vitreous
body of the eye \cite{lee} have been measured. It is also possible to
do microrheological measurements by a quantitative measurement of the 
mean square displacement $<\Delta r^{2}(t)>$ of the probe particles 
caused due to thermal fluctuations. This can be done either using laser
interferometry with a resolution less than 1 nm \cite{git,sch} or by 
diffusing wave spectroscopy\cite{mas}.    
 
The motion of the probe particle of radius 'a' may be described by a 
generalized Langevin equation given by :
\begin{equation}
m\dot v(t)=f_{R}(t)-\bigl\lmoustache _{0}^{\infty}\zeta (t-\tau )v(\tau 
)d\tau 
\end{equation} 
where $m\dot v(t)$ is the inertia of the particle, $f_R(t)$ is   
the contribution due to electrostatic and Brownian forces on the 
particle. $\zeta$ defines a time-dependent memory function 
which contributes to the viscous damping of the particle in the 
viscoelastic medium. The memory function $\zeta(t)$ and $f_R(t)$ 
are related by the following temporal autocorrelation function:
\begin{equation}
<f_R(0)f_R(t)>=k_B T \zeta (t)
\end{equation}
where $k_B$ is the Boltzmann's constant and T is the temperature.

In the frequency domain, the viscosity of the medium may be 
related to the frequency dependent memory function by the 
generalized Stokes-Einstein relation 
\begin{equation}
\widetilde \eta (s)={\widetilde \zeta (s)\over 6\pi a} 
\end{equation}
where s is the complex frequency given by $s=\it i \omega$. The  
complex viscoelastic modulus is given by \cite{mas}
\begin{equation}
\widetilde G(s)=s\widetilde \eta (s)={s\over 6\pi a}[{6k_{B}T\over s^{2}<\Delta
r^{2}(s)>}-ms]  
\end{equation}
The last term on the right hand side of Equation 11 is the
contribution due to the inertia of the particle and can be 
omitted except at very high frequencies. $<\Delta r^2 (t)>$ of 
the 
probe particle is obtained from the intensity                 
autocorrelation function which can be measured in diffusing wave 
spectroscopy experiments in the transmission or backscattering 
geometries. The Laplace transform of $<\Delta r^2 (t)>$ is 
used to calculate $\widetilde G(s)$ using Equation 11.
$\widetilde G(s)$ 
is then fitted to a functional form in s,   
which may then be used to calculate $G^\star (\omega)$, using 
the 
method of analytic continuation. 

We have used microrheology to estimate the $G^\prime(\omega)$ 
and 
$G^{\prime\prime}(\omega)$ 
of an aqueous solution of CTAT of weight fraction 1\%. The probe 
particles used are polystyrene colloidal particles of 
diameter 0.23 $\mu$m dispersed in water at $\phi=1\%$. Diffusing  
wave spectroscopy was performed on the equilibrated sample 
using our light scattering setup consisting of a  
$Kr^+$ ion laser (model 2020, Spectra Physics, U.S.A.,  
excitation wavelength used 647.1 nm), a homemade 
spectrometer, photomultiplier tube (model R943-02, Hamamatsu, 
Japan), single photon amplifier discriminator (SPEX) 
and a MALVERN 7132 
CE 64 channel correlator (Fig. 7). The light scattered I$_{s}$(t)
by the probe 
particles in the backscattering direction at a temperature of 
25$^\circ$C is used to measure
the normalized intensity autocorrelation 
function $g_2(t) = <I_{s}(0)I_{s}(t)>/<I_{s}(0)>^{2}$, as shown in 
Fig. 8 (a). For backscattering geometry,
$g_{2}(t)\sim e^{-\Gamma(k^{2}<\Delta r^{2} (t)>)^{1/2}}$ 
\cite{pin}, which is used to get 
$<\Delta r^{2}(t)>$ as shown in Fig 8 (b), using 
$\Gamma$ = 2. The parameter $\Gamma$ is a constant 
depending on the polarisation of the scattered light and varies inversely 
with the transport mean free path l$^{\star}$ of the diffusing photon. 
Fig 8 (c) shows G(s) 
calculated using Equation 11, which was fitted to
$G(s) = p_{o} + p_{1}s^{-0.55} + p_{2} s^{0.3} + p_{3} s^{0.5} + p_{4}s$
\cite{mas}.
Putting $s = i\omega$, $G^\prime(\omega)$ and 
$G^{\prime\prime}(\omega)$ of the dispersing gel are
 calculated as shown in 
Fig. 8(d). Comparison of Figs. 4 and 8 (d)
shows similar magnitudes of the viscoelastic response 
functions obtained by macrorheology and 
microrheology methods. Further, in Fig. 8(d),
the crossover of 
the viscous and elastic moduli occur at
$\omega _{co}\sim $0.4 rad/sec, indicating a relaxation time
$\tau_{R} \sim$ 2.5 seconds for 1\% CTAT at $25^\circ$C, similar to 
macrorheology measurements.
It may be noted that
microrheology may be used to calculate 
the frequency response of CTAT to much higher frequencies than 
the conventional rheometer experiment. 

\section{Nonlinear rheology of CTAT}

To study the nonlinear rheology of CTAT, we have measured the 
flow curve of CTAT 1.35wt.\% at 25$^\circ$ C as shown
in Fig. 9 . The measurements are 
done under conditions of controlled stress. The data points 
are 
collected at intervals of 1 second, a value comparable to the 
relaxation time of the sample. The resultant branch of the   
measured flow curve is  
metastable, and its existence was demonstrated by 
Grand 
et al\cite{gra}. The flow curve is found to saturate to a 
constant stress value above a critical shear rate $\dot\gamma_c$,  
while the first normal stress difference is found to increase 
linearly with shear rate. The plateau of the shear stress at 
high shear rates in CPyCl/NaSal has been interpreted by Grand et 
al\cite{gra} as a characteristic feature of the flow curves of 
complex fluids that gives rise to a mechanical instability of 
the 
nature of shear banding\cite{spe}. Shear banding results in 
the formation 
of bands of high and low viscosities in the sample, supporting 
low and high shear rates, respectively. However, the 
same phenomenon observed in CTAB/NaSal at a higher              
concentration has been explained by 
Berret et al\cite{ber} as due to the coexistence of isotropic and 
nematic phases in the sheared sample. 

In addition to the measurement of the flow curve for our system, 
we have studied the stress relaxation in the sample after 
subjecting it to a step strain rate. At 25$^\circ$C, on  
applying controlled shear rates whose values  
lie in the plateau region of the 
flow curve, the stress, instead of decaying to a steady       
state, is found to oscillate in time. 
 Fig. 10 shows the time dependent stress relaxation in
the 1.35wt.\% CTAT sample at 25$^\circ$C, on subjecting the
sample to a step strain rate of 100s$^{-1}$.
The Fourier spectra of these time-dependent signals show 
time scales of the order of a few tens of seconds, which are an 
order of magnitude larger than $\tau_{R}$. 

We identify the observed time-dependent behaviour 
as a manifestation of the mechanical instability
due to the formation of shear bands\cite{spe,gra}. 
Preliminary 
analysis of the time series obtained from the stress relaxation 
experiments done in the nonlinear regime shows  
the existence of positive Lyapunov exponents 
\cite{gao} and finite 
correlation dimensions \cite{gra1}($>$ 2 at shear rates $>$ 75 s$^{-1}$), 
which points to the 
existence of deterministic chaos in sheared aqueous solutions of 
CTAT. The Lyapunov exponent characterizes the divergence of 
stress
trajectories in the system, whereas the correlation dimension gives us
information about the geometry of the attractor on which the 
trajectories in phase space asymptotically lie. 
On increasing the temperature of the sample to             
35$^\circ$C, and on maintaining the same shear rates as in the 
previous experiments done at 25$^\circ$C,  
the time dependent oscillations in the stress 
relaxation are found to disappear completely. 
This is in accordance 
with previous studies on the temperature dependence of the 
flow 
curve of CPyCl/NaSal \cite{ber} which shows a decrease in the width 
of the 
plateau with increasing temperature. The disappearance of the 
time-dependent behaviour in sheared CTAT at higher temperatures 
is thus a direct consequence of the disappearance of the shear 
bands in the sample\cite{ber}. We have done extensive studies on the 
time-dependence of the stress relaxation of dilute, aqueous,  
sheared solutions of CTAT by doing more elaborate analysis of 
the time-series obtained from our experiments\cite{ran}.

AKS thanks Board of Research in Nuclear Sciences and 
RB thanks the CSIR for financial support. We thank Sriram Ramaswamy, 
P. R. Nott and V. Kumaran for the use of the rheometer. We  would like to
acknowledge  M. E. Cates' lectures on
rheology delivered at the Summer School on Soft Condensed Matter, International
Center of Theoretical Physics, Italy, May 4 - June 5, 1998.

\begin{figure}
\caption{(a) shows step strain applied at time $t_1$, with a magnitude
small enough to lie in the linear regime. (b)
shows the corresponding relaxation of stress.}
\end{figure}

\begin{figure}
\caption{ $\gamma$ is the shear applied to a viscoelastic material. 
$\sigma$ is the resultant stress in the material, delayed by a phase
angle $\delta$. The complex shear modulus may be then written as
$G^{\star}(\omega) = G^{\prime}(\omega)+i G^{\prime\prime}(\omega)$, where
 $G^{\prime}(\omega)$ is the in-phase elastic modulus and 
$G^{\prime\prime}(\omega)$ is the out-of-phase
 viscous modulus.}
\end{figure}

\begin{figure}
\caption{(a) shows Newtonian ($\sigma_{xy} = \eta \gamma_{o}$), 
shear 
thinning and shear thickening behaviours seen in viscoelastic fluids.
(b) shows the flow curve of a Herschel-Bulkley plastic with a finite yield stress
$\sigma_{o}$. At $\sigma_{xy} < \sigma_{o}$, the material behaves like a solid.
At $\sigma_{xy} > \sigma_{o}$, 
$\sigma_{xy} = \sigma_{o} + K_{p}\dot\gamma ^{p}$, 
which reduces to 
the Bingham equation for p=1.
(c) shows the flow curve of a system undergoing a flow-induced phase 
transition, characterised by a sudden jump in $\sigma$ as shown 
by the dotted line.
(d) shows a flow curve showing a plateau region, which is a signature of 
shear banding.}
\end{figure}

\begin{figure}
\caption{ Measurement of the elastic modulus $G^{\prime}(\omega)$ and the
viscous modulus $G^{\prime\prime}(\omega)$ of 1wt.\% CTAT by the method of
macrorheology, using a cone and plate rheometer, with applied stress = 0.08 Pa
at 25$^\circ$C.}
\end{figure}

\begin{figure}
\caption{ (a)  $G^{\prime}(\omega)$ and (b) $G^{\prime\prime}(\omega)$ of 
1wt.\% CTAT 
using a rheometer (data same as shown in Fig. 4). The solid lines show the 
least square fits to
the Maxwell model. (c) shows the Cole-Cole plot which deviates from the 
semicircle exhibited by Maxwellian systems.}
\end{figure}

\begin{figure}
\caption{ Data same as in Fig. 4  and the corresponding fits to the
Doi-Edward model shown by solid lines.}
\end{figure}

\begin{figure}
\caption{ Our light scattering setup in the transmission geometry : P1 and P2 are the analyser and
polariser respectively, L1 and L2 are convex lenses of f = 20 and
30 cm respectively, P.H. is a pin hole, S the sample that scatters light
and Inv is an inverter circuit. M1 and M2 are plane mirrors that steer the incident laser
beam.}
\end{figure}

\begin{figure}
\caption{ The microrheology results:
(a) shows the correlation function in the backscattering
direction, (b) the $<\Delta r^{2}(t)>$, (c) the complex modulus G(s) 
and (d) shows the calculated
values
of $G^{\prime}(\omega)$ and $G^{\prime\prime}(\omega)$.}
\end{figure}

\begin{figure}
\caption{ The metastable branch of the
flow curve of 1.35wt.\% CTAT, measured under conditions of
controlled stress .}
\end{figure}

\begin{figure}
\caption{The time-dependent relaxation of stress in
1.35wt.\% CTAT, on
subjecting the sample to a constant step-strain rate of 100 s$^{-1}$.}
\end{figure}

 \begin{figure}
\centerline{\epsfxsize = 8cm  \epsfbox{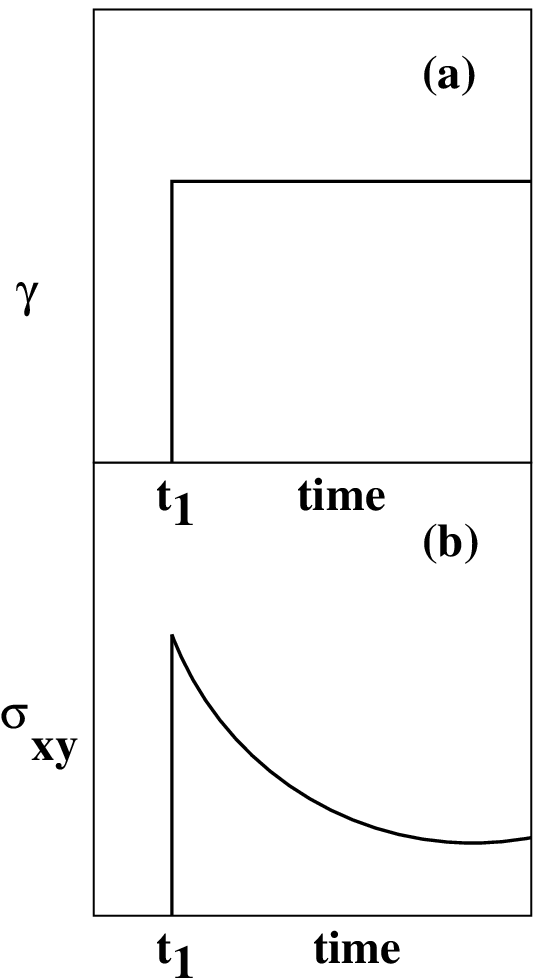}}

\label{Fig.1}

\end{figure}
\flushbottom{\bf{Fig. 1}}

\begin{figure}
\centerline{\epsfxsize = 8cm \epsfbox{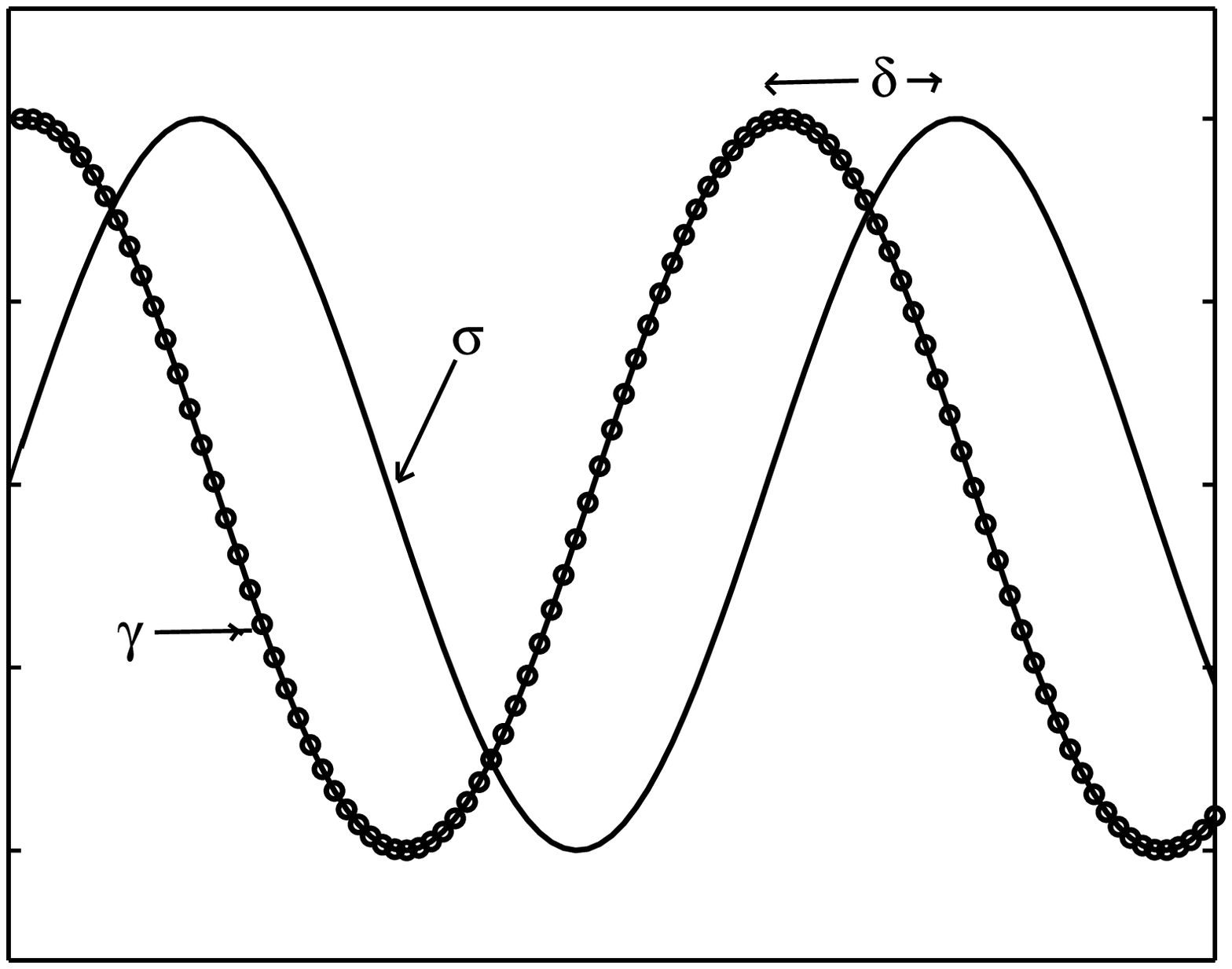}}

\label{Fig.2}
\end{figure}
\flushbottom{\bf{Fig. 2}}

\begin{figure}
\centerline{\epsfxsize = 8cm \epsfbox{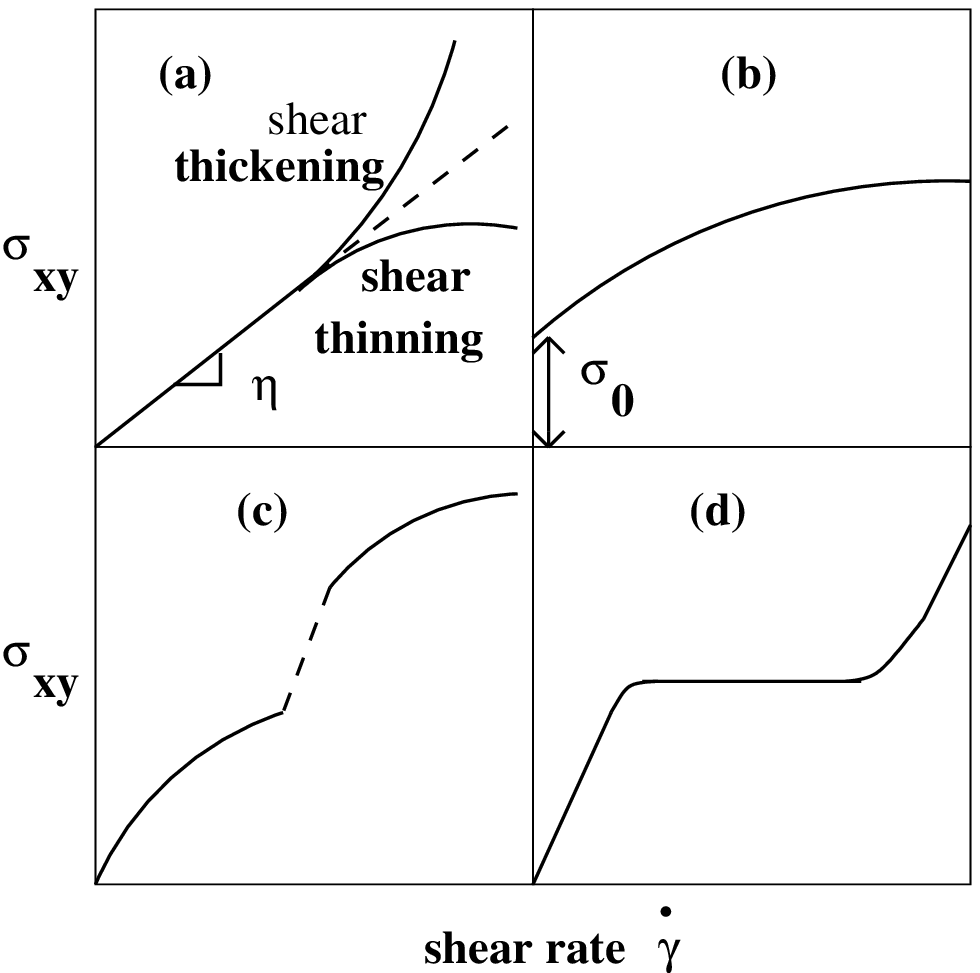}}

\label{Fig.3}
\end{figure}
\flushbottom{\bf{Fig. 3}}

\begin{figure}
\centerline{\epsfxsize = 8cm i\epsfysize = 8cm \epsfbox{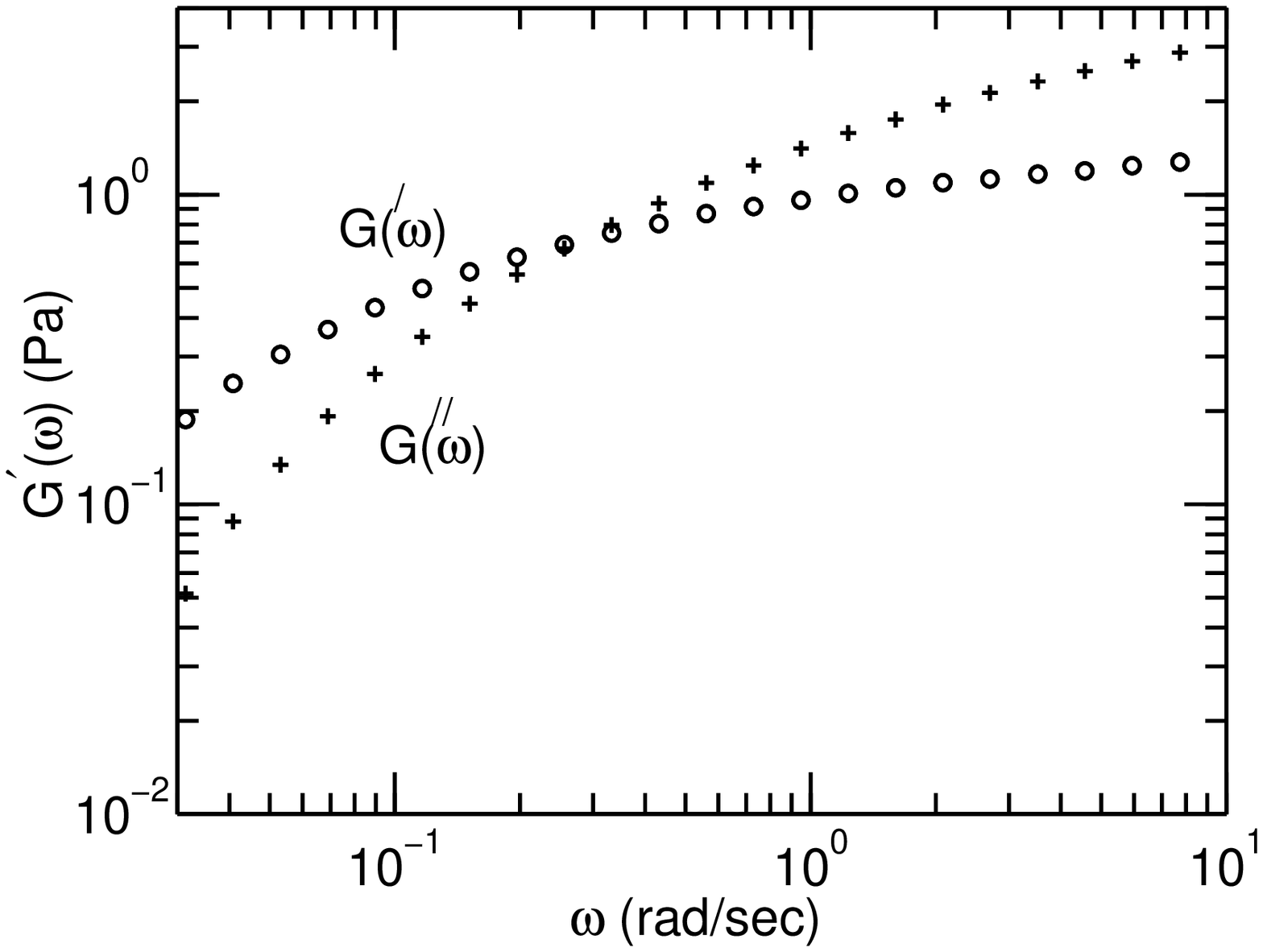}}
\label{Fig.4}
\end{figure}
\flushbottom{\bf{Fig. 4}}

\newpage
\begin{figure}
\centerline{\epsfxsize = 8cm \epsfbox{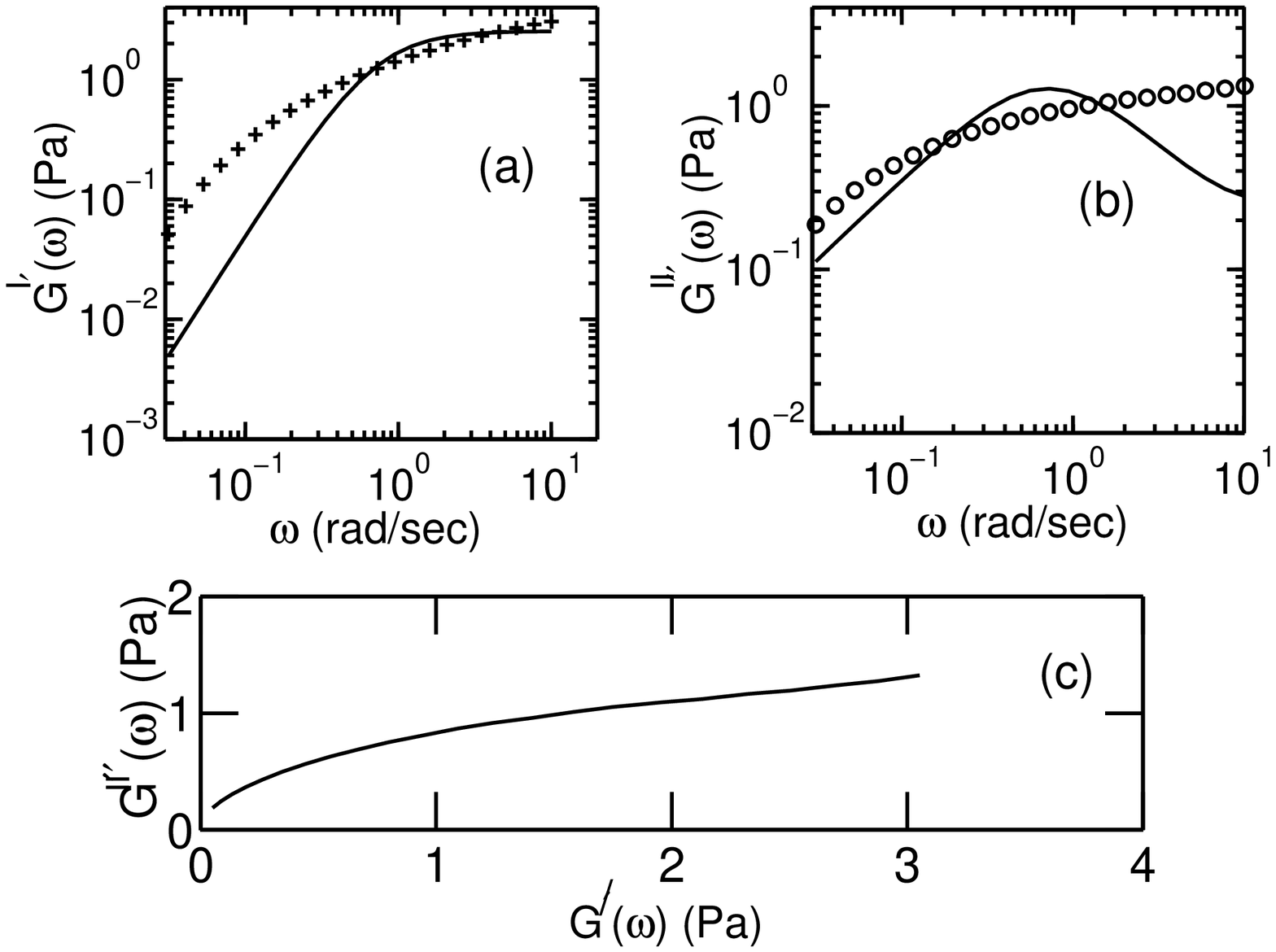}}

\label{Fig.5}
\end{figure}
\flushbottom{\bf{Fig. 5}}

\begin{figure}
\centerline{\epsfxsize = 8cm \epsfbox{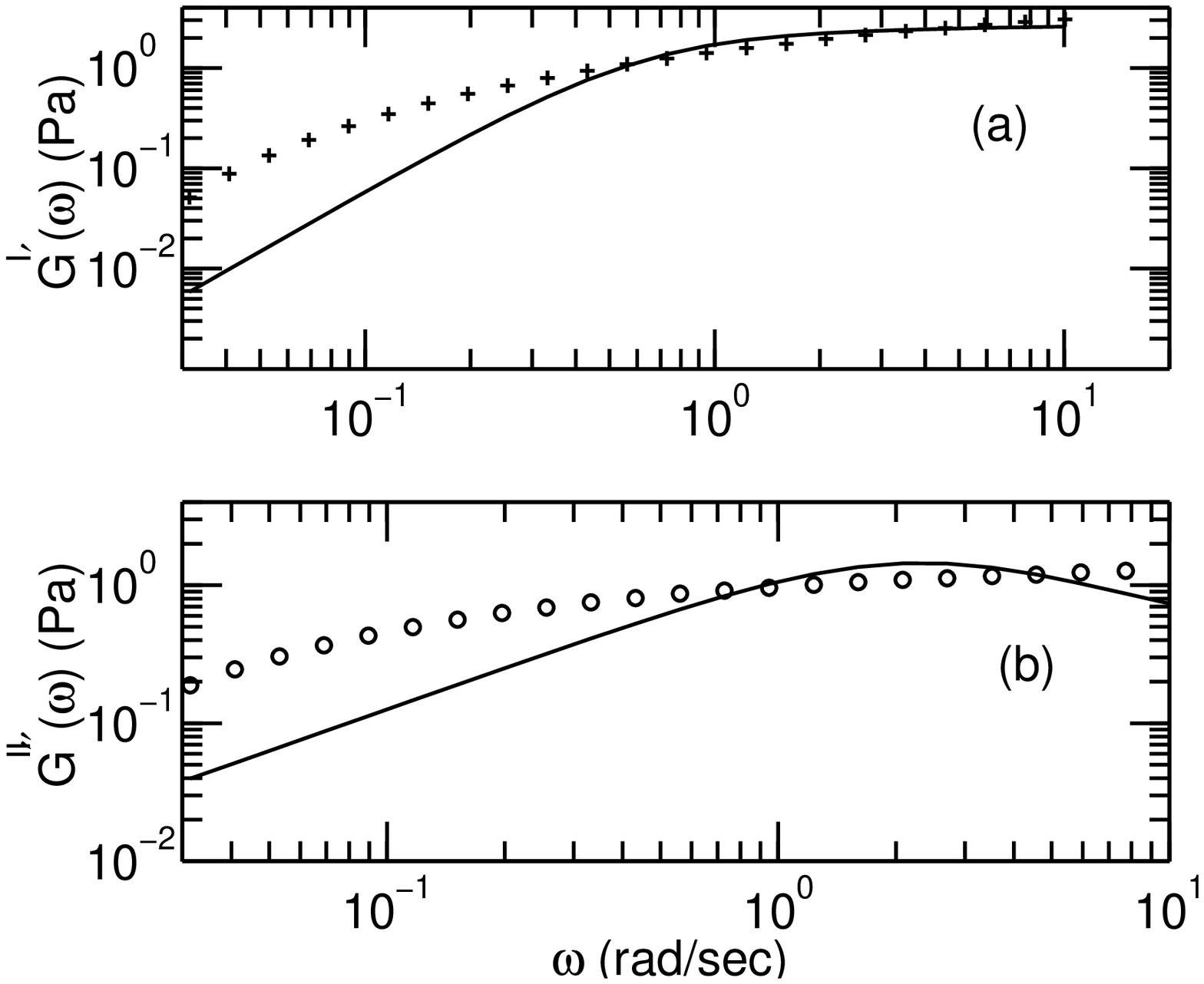}}
\label{Fig. 6}
\end{figure}
\flushbottom{\bf{Fig. 6}}

\newpage
\begin{figure}
\centerline{\epsfxsize = 8cm, \epsfysize = 8cm \epsfbox{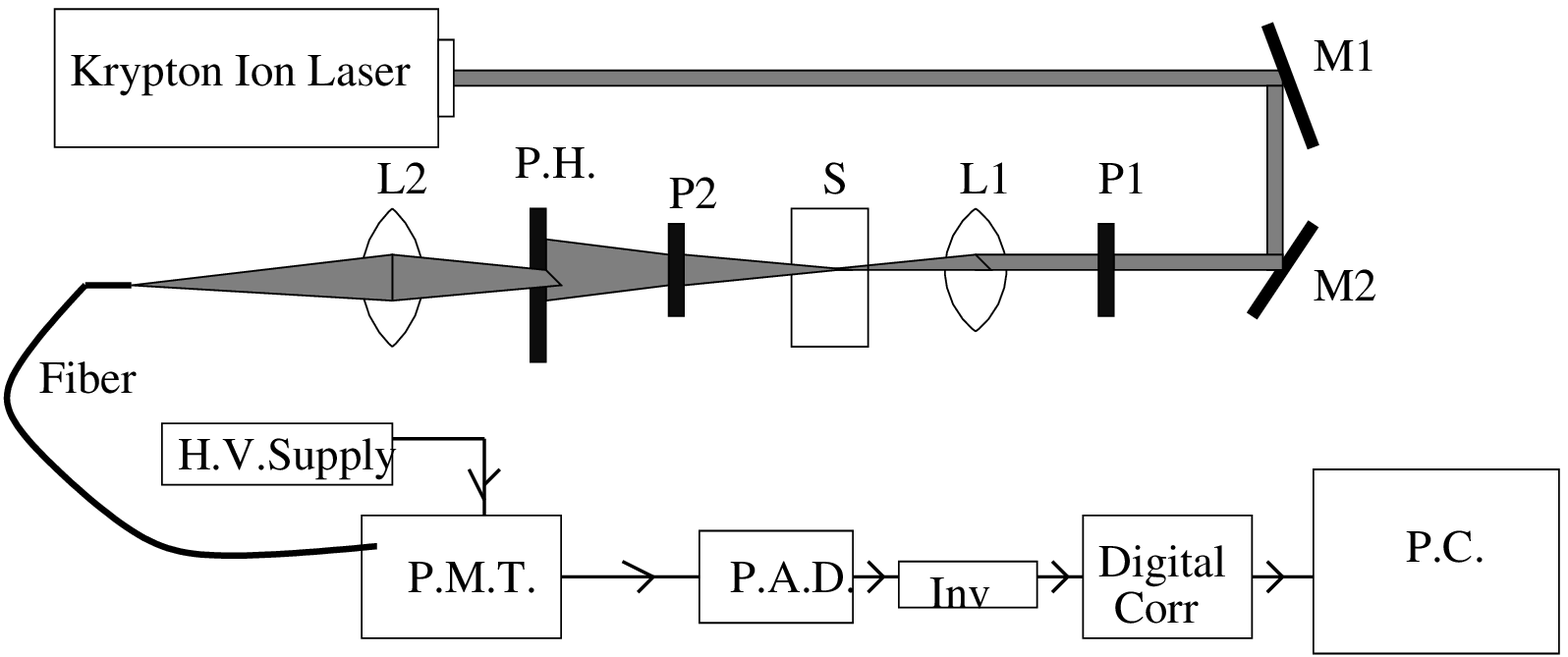}}
\label{Fig.7}
\end{figure}
\flushbottom{\bf{Fig. 7}}

\begin{figure}
\centerline{\epsfxsize = 8cm \epsfbox{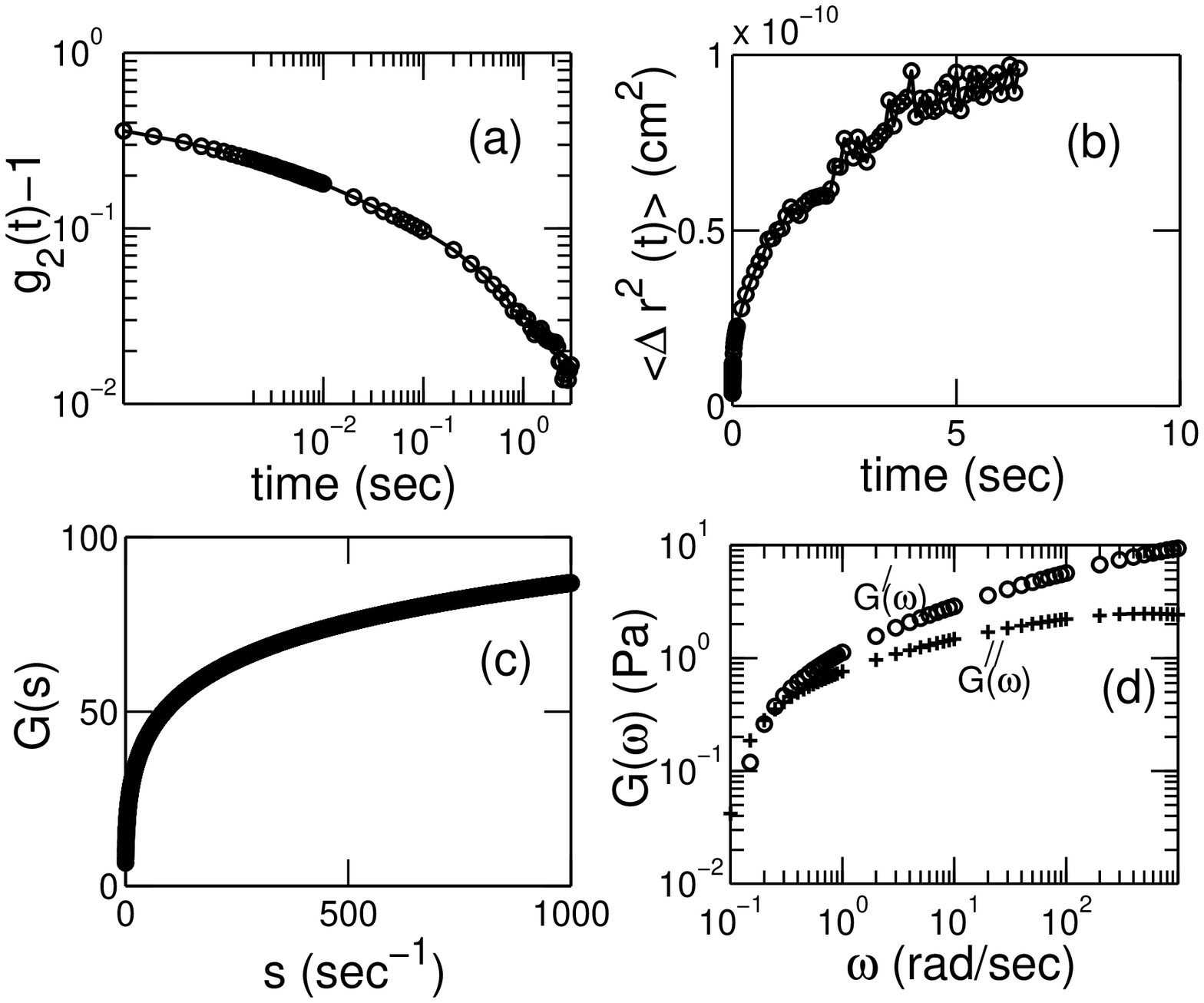}}
\label{Fig. 8}
\end{figure}
\flushbottom{\bf{Fig. 8}}

\begin{figure}
\centerline{\epsfxsize = 8cm \epsfbox{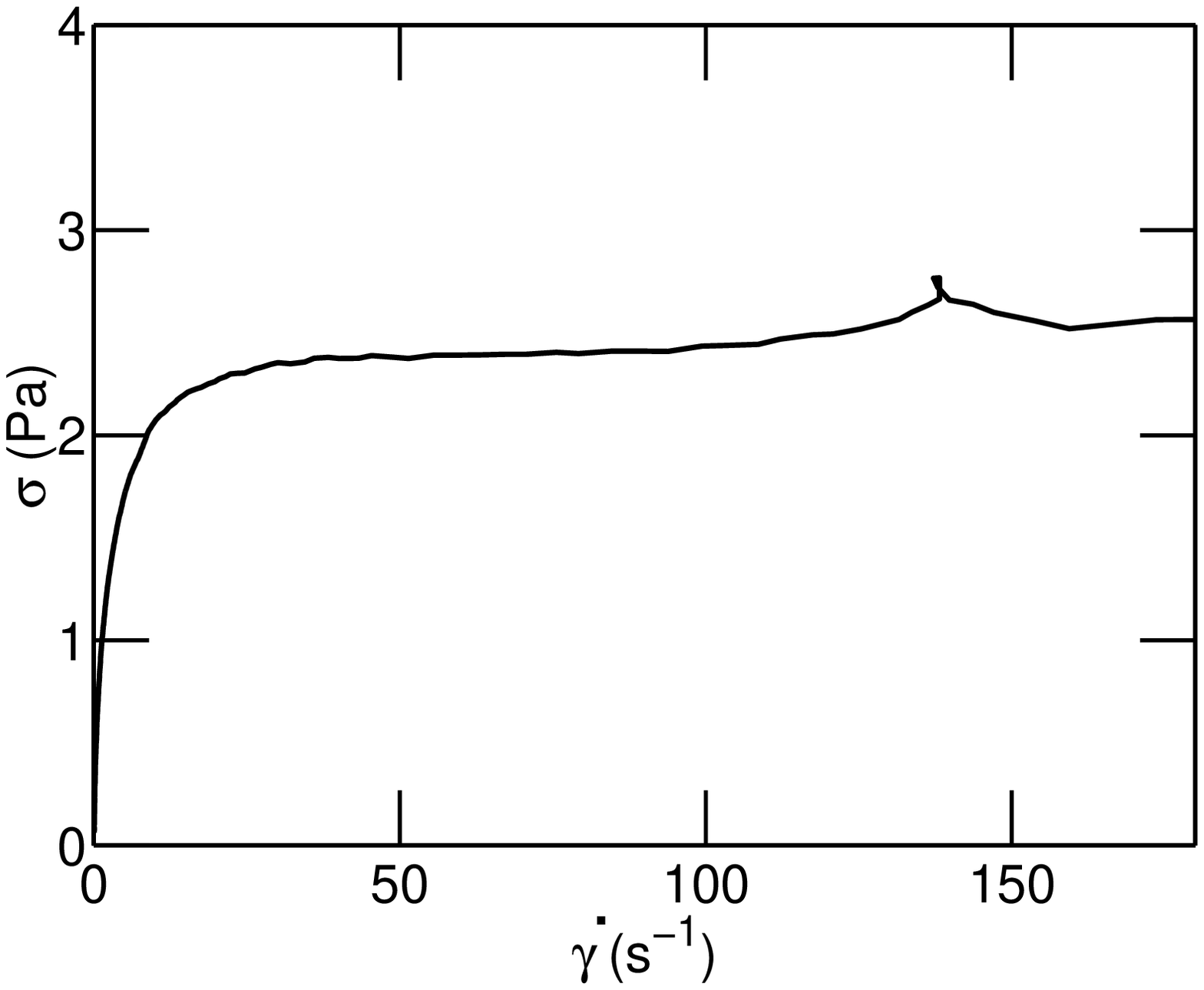}}

\label{Fig.9}
\end{figure}
\flushbottom{\bf{Fig. 9}}

\begin{figure}
\centerline{\epsfxsize = 8cm \epsfbox{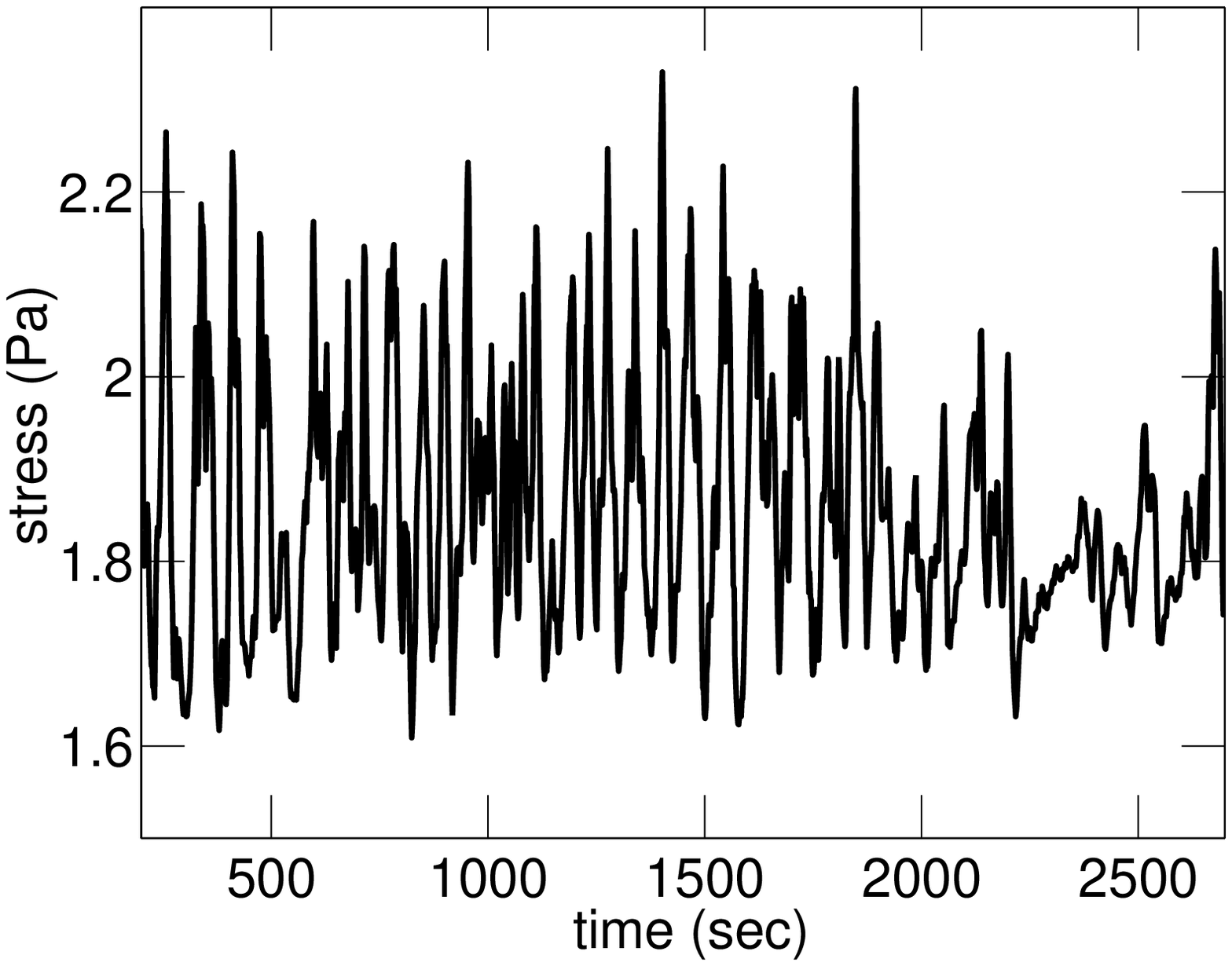}}

\label{Fig.10}
\end{figure}
\flushbottom{\bf{Fig. 10}}

\end {document}